\documentclass[fleqn,twoside]{article}
\usepackage{espcrc2,psfig}

\usepackage{graphicx}
\usepackage[figuresright]{rotating}

\title{
Searching for high-redshift
centimeter-wave continuum,
line and maser emission using the Square Kilometer Array} 

\author{A.\,W. Blain\address{California Institute of Technology, Pasadena, CA,
USA, awb@astro.caltech.edu}, 
C. Carilli\address{NRAO, Socorro, NM, USA, ccarilli@nrao.edu},
J. Darling\address{OCIW, Santa Barbara Street, Pasadena, CA, USA, 
darling@ociw.edu}
}

\begin{document}

\begin{abstract}
We discuss the detection of redshifted line and continuum emission at radio 
wavelengths using a Square Kilometer Array (SKA), specifically 
from 
low-excitation rotational molecular line transitions of CO and HCN 
(molecular lines), the 
recombination radiation from atomic transitions in almost-ionized hydrogen  
(radio recombination lines; RRLs), OH and H$_2$O maser lines, as well as 
from synchrotron and free-free 
continuum radiation and H{\sc i} 21-cm line radiation. 
The detection 
of radio lines with the SKA offers the prospect
to determine the redshifts and thus exact luminosities for some of the 
most distant and optically faint
star-forming galaxies and active galactic nuclei (AGN), 
even those galaxies that are either 
deeply enshrouded in interstellar dust or shining prior to the 
end of reionization. Moreover, it 
provides an opportunity to study the astrophysical conditions and 
resolved morphologies of the most active regions 
in galaxies during the most active phase of star formation at 
redshift $z \sim 2$. A sufficiently powerful and adaptable SKA correlator 
will enable wide-field three-dimensional 
redshift surveys at chosen specific high redshifts, and 
will allow new probes of the 
evolution of large-scale structure (LSS) in the distribution of galaxies. 
The detection of molecular line radiation favours pushing the 
operating frequencies of SKA up to at least 26\,GHz, and ideally to  
40\,GHz, while very high redshift maser emissions requires access 
to about 100\,MHz. To search for LSS
the widest possible instantaneous field of view 
would be advantageous. 
\end{abstract}


\maketitle

\section{Introduction}

The processes that established the properties of the populations of galaxies 
we see in the Universe today have begun to be 
probed in detail only during the 
last decade (see summaries at long and short wavelengths 
by Smail  et al.  2002 and
Steidel  et al.  2003 respectively). 
Information about the evolution of galaxies out to redshifts $z \sim 5$ and 
beyond is now being gathered as an ever increasing rate. Multiwavelength 
studies have been confirmed as vital to investigate the nature of the 
processes at work 
in forming galaxies (e.g. Alexander  et al.  2003), and radio 
observations have played a major part in these developments (e.g.
Chapman  et al.  2003b). However, 
existing radio telescopes are only sensitive enough to detect the 
most luminous galaxies (brighter than 
at least a few $10^{12}$\,L$_\odot$) at 
redshifts greater than two. A much more sensitive radio instrument would 
provide a better match to the capabilities of the 
{\it James Webb Space Telescope 
(JWST)}\footnote{http://ngst.gsfc.nasa.gov/}, the Atacama 
Large Millimetre Array (ALMA)\footnote{http://www.alma.nrao.edu/} 
and future 30-m-class ground-based 
telescopes such as the TMT under study.\footnote{http://www.astro.caltech.edu/observatories/tmt/}  

Key questions that remain in galaxy formation include the evolution 
of the luminosity function of galaxies prior to the apparent peak of 
starformation activity at $z \simeq 2-3$ (Madau  et al.  1996; Blain  et al.  
1998; Chapman  et al.  2003a; Giavalisco  et al.  2004; Steidel  et al.  2004). 
The maximum luminosity of galaxies at 
such early epochs is still little known, although very luminous 
examples have been found (Becker  et al.  2001; Fan  et al.  2004). 
The evolution of the 
space densities and luminosities of galaxies  
extending through the newly established 
epoch from the end of reionization at $z \simeq 6.5$, 
to its beginning 
at $z \simeq 20$, as highlighted by the recently detected Gunn--Peterson 
effect and cosmic microwave background (CMB) polarization  
observations respectively 
(Spergel  et al.  2003) is entirely unresolved by existing 
observations. 

Because reionization has almost no effect on 
the opacity of the Universe to radio waves, a very sensitive radio telescope 
could be tremendously useful for finding and 
revealing the processes at work in the  
galaxies that reionized the intergalactic medium (IGM) at 
$z>6$. The role of feedback from star formation and AGN 
activity in modifying the stellar formation rate, populations and 
AGN accretion is also 
little known, but must be sensitive to the masses and densities of 
galaxies, that can be probed using radio line observations deep into 
dust obscured regions. 
Radio continuum emission can provide insight 
into the power of supernovae and the density of the interstellar 
medium (ISM) in these early objects.  The low opacities of both the 
IGM and ISM to radio 
and mm-wave line emission ensures that this radiation 
could provide information on 
the masses and excitation conditions right into the cores of the very  
earliest galaxies that formed. 

Alongside studies of the properties of individual forming galaxies, the 
richness of large-scale structure (LSS) in the three-dimensional 
distribution of galaxies has been 
resolved in unprecedented 
detail, both at moderate redshifts (Blanton  et al.  2003), and out to 
greater distances (Venemans  et al.  2002; Adelberger  et al.  2003; 
Blain  et al.  2004; Coil  et al.  2004). 
Overdensities in the distribution 
of galaxies found near known extremely 
luminous objects have also been resolved (Brand  et al.  2003; 
Stevens et al.\ 2004). 
In the future, radio spectroscopy could have a key role to play in resolving 
the three-dimensional structures traced by very large numbers of normal 
galaxies over wide fields at high redshifts, building on redshift 
determinations now possible using the Westerbork WSRT\footnote{
http://www.astron.nl/p/observing.htm} and the Greenbank 
Telescope (GBT).\footnote{http://www.gb.nrao.edu/GBT/GBT.shtml}  
Radio spectroscopy has the advantage of tracing the bulk of 
the ISM in galaxies, and so accurate relative 
velocities between nearby galaxies 
can be obtained, without concern for the effects of 
winds and outflows to confuse relative velocities (Shapley  et al.  2003), 
and blur out 
velocity space distortions that reveal the relaxation of clusters of 
galaxies. 

\subsection{Existing radio observations} 

The faintest galaxies that can be detected 
detected in the radio continuum using the NRAO VLA 
have been studied for almost 
20 years (Donnelly, Partridge \& Windhorst 1987), and the 
$\mu$Jy population continues to promote 
important insight into the galaxy formation process to the 
present day, providing an extinction independent, high-resolution view 
of the evolution of galaxies (Haarsma  et al.  2000; 
Chapman  et al.  2003b; Cowie  et al.  2004). However,  
spectroscopy at cm wavelengths using existing 
instruments remains difficult, owing to very 
restrictive correlator 
bandwidths, and only single very luminous objects can be 
targeted for radio spectroscopy at the VLA 
(e.g. Walter  et al.  2003). 

The availability of radio observations has 
proved to be very valuable in the interpretation of the nature of 
new populations of dust-enshrouded 
galaxies that are relatively difficult to 
detect in all but the deepest images at optical and wavelengths 
(Smail  et al.  1997, 2002; Frayer  et al.  2004), 
but nevertheless have luminous dust 
continuum emission at mid- and far-infrared(IR) and submillimeter (submm) 
wavelengths 
(Altieri  et al.  1999; Smail, Ivison \& Blain 1997; Smail  et al.  2002; 
Bertoldi  et al.  2002, 2004; Greve  et al.  2004). These galaxies
that contribute a significant fraction of the 
luminosity of all galaxies at  
high redshifts (Blain  et al.  1999). Radio observations have been crucial in pinpointing the 
positions of these galaxies 
with sufficient accuracy to enable optical spectroscopy, changing them 
from an interesting but anecdotal population of objects with uncertain 
properties into a well studied sample of galaxies  
(Chapman  et al.  2002, 2003a, 2004; Smail  et al.  2003; Blain  et al.  2004). 

All observations of distant galaxies in the 
radio waveband benefit from negligible extinction, and  
well-designed interferometers offer both 
naturally fine angular resolution and wide instantaneous fields of view. 
With sufficient collecting area the instrumental sensitivities  
are excellent (at the $\mu$Jy level), and the key 
requirement for spectroscopy is a sufficiently powerful correlator 
to provide the bandwidth to conduct redshift surveys across 
the whole degree-scale instantaneous 
field of view.\footnote{A 12-m scale element at 1\,GHz 
provides a 1.5-deg-wide field of view, while a 100-km baseline 
corresponds to a resolution of 0.06\,arcsec.} This class of 
performance would be tremendously powerful for probing high-redshift 
LSS, as the largest scales of order 100\,Mpc 
manifested in the galaxy distribution correspond 
to about 3-4\,deg on the sky at high redshifts, while 
individual distant galaxies 
subtend angles of order 0.5\,arcsec. Resolved images and three-dimensional 
spectra of all the 
members of a forming high-redshift cluster 
of galaxies could thus be obtained in 
only a few pointings of SKA. 

The flexibility and power of a correlator capable of wide-field 
spectroscopy could also be very adaptable to multi-beam-forming 
observations of widely-separated 
known targets or for continuum surveys over several
fields of view. 

The prospects for detecting the 
radio synchrotron and free--free continuum emission from distant, 
low-luminosity galaxies are excellent -- while existing radio surveys 
reach down to of order 10\,$\mu$Jy at 1.4-GHz, the proposed 
sensitivity of the SKA corresponds to a depth of 80\,nJy RMS 
in 24\,hr over a 
field that could be as wide as 1.2\,deg$^2$. Arcsec-scale resolution 
will be essential to avoid confusion at the faintest levels. Note that the 
detection rate of galaxies at redshifts greater than unity of about 
$10^4$\,hr$^{-1}$ is unprecedented in 
astronomy. 

A huge task remains to find redshifts for these faint radio galaxies, 
many of which, and perhaps some of the most interesting examples,  may not be 
easily accessible to optical or near-IR 
spectroscopy, either because of a lack of 
emission lines,  
or to extreme distance and faintness. Fine angular resolution ensures that 
the problem is not as severe for SKA-detected galaxies as for those 
selected at far-IR or submm 
wavelengths. The discovery images at shorter wavelengths have 
relatively poor angular 
resolution, including the 
forthcoming avalanche of many millions of new galaxies expected to be 
discovered by {\it Spitzer Space Telescope 
(Spitzer)}\footnote{http://www.spitzer.caltech.edu/}  
following its August 2003 launch (Werner, Gallagher \& Irace, 2004), 
and the fainter galaxies to be detected 
using the 
future 3.5-m {\it Herschel Space 
Observatory} after 2008\footnote{http://herschel.jpl.nasa.gov/}. 
In part, ground-based integral field unit spectrographs 
can locate galaxies when positions are known less accurately than a 
few arcsec, but to compile large samples and substantially 
complete redshift catalogs, 
techniques that rely on neither accurate 
positions nor optical telescopes will probably be required. Several 
types of diagnostic lines are available to 
be probed at radio wavelengths, both the 
centimeter and millimeter bands, and these 
could offer a breakthrough with future radio telescopes. 
The SKA could be the workhorse instrument to unlock 
the multiwavelength Universe, 
much as the VLA has been in the last decade for the 
existing small samples of the most luminous high-redshift 
galaxies (Smail  et al.  2002; Chapman  et al.  2003a,b, 2004). 

The advantage that a radio interferometer, with appropriately small 
individual elements has over the ALMA mm/submm-wave interferometer is 
a potentially much greater instantaneous 
field of view, and a greater chance of serendipitous detection 
of line emission at very high redshifts (Carilli \& Blain 2002). 
ALMA can carry out the necessary 
observations in molecular lines (Blain et al. 2000), 
but only for a relatively tiny instantaneous 
field of view. Operating at much longer wavelengths, a radio array 
can cover a much wider field, while using longer baselines to maintain angular 
resolution, detecting molecular transitions, recombination and maser 
lines in 
relatively low excitation gas. 

\section{Line emission mechanisms} 

\subsection{Low-excitation molecular lines}

The fuel for star formation in both the local and distant 
Universe is dense molecular gas,
cooled by optically thin emission from molecular and metal fine-structure 
lines, mechanisms which require prior enrichment of the ISM 
in the galaxies. In the 
earliest objects molecular hydrogen would be the cooling mechanism 
(Rees \& Ostriker 1977; Sokasian  et al.  2004). 
Hence, the second generation of 
star formation, potentially 
deep in the epoch of re-ionization at $z \simeq 6-15$ should generate 
powerful 115-GHz CO(1-0) line emission, while the densest regions could 
excite and emit 88.6-GHz HCN(1-0) radiation.   

Working in the radio waveband, rotation transitions between only the 
lower-$J$ states of polar molecules are available for observations. 
The most promising molecules are CO and HCN (from which emission 
is likely to be an order 
of magnitude fainter than CO). Other molecules like HCO$^+$ might 
be as promising as HCN for high-redshift 
studies. 
Depending on the maximum frequency of the SKA bands, CO and HCN could 
be observed for all redshifts greater than about 1.6 and 0.95 
respectively (for an upper frequency limit of 44\,GHz matched to the 
limit of atmospheric transparency). For 
very high-redshifts ($z>9.5$), should metal-enriched 
galaxies exist there, higher-excitation lines 
up to CO(4-3) might also be detectable, redshifted into this potential 
operating range of the SKA extending to 
about 40\,GHz.
At such very high redshifts, 
the increased CMB temperature ($\simeq 30$\,K) 
would allow higher-$J$ level transitions to 
occur, as the relevant rotational levels of CO and HCN will be 
excited. Note, however, that the 
lines will not be made more luminous 
by scattering off higher-energy 
CMB photons. 

Molecular line emission from the 115-GHz CO(1-0) transition has been 
detected from one of the most distant QSOs currently known (Walter  et al.  
2003), and from other very luminous, high-redshift galaxies (Carilli  et al.  
2002, 2004; Lewis  et al.  2002; Greve  et al.  
2003; Solomon  et al.  2003).
The 10 to few-100\,km\,s$^{-1}$ 
width of detected lines is likely to 
provide a good estimate of the velocity dispersion of the gas 
in the system in which it is detected (Sheth et al.\ 2004; 
Tacconi et al. 2005), 
and if sub-arcsec resolution 
is available, then an accurate mass can be derived for organized 
systems.

Out to $z \simeq 20$ at the likely start of reionization 
(Spergel  et al.  2003), CO(1-0) 
emission could be detected at a frequency of 5--8\,GHz, if sufficiently 
luminous galaxies were present. This is a speculative, but not 
impossible goal for a future generation of very sensitive radio telescopes. 

Fig.\,2 presents the number of CO, fine-structure and 
radio recombination lines expected at 4 chosen 
example 4-GHz wide SKA observing bands. The calculations follow those 
outlined in Carilli \& Blain (2002) and Blain  et al.  (2000). 
The results have been updated to include relatively small 
shifts in the 
fitting parameters required by the new redshift distribution for 
submm/radio-selected galaxies determined by 
Chapman  et al.  (2003a; 2004), 
and to take account of the post-{\it WMAP} cosmological parameters
($\Omega_{\rm m}=0.27$, $\Omega_{\Lambda}=0.73$ and $H_0=71$\,km\,s$^{-1}$): 
see Knudsen et al.\ (2005). 

\subsection*{Galaxies in absorption: CO and H{\sc i}} 

With excellent resolution in both the spectral and spatial domains, 
SKA absorption line spectroscopy should be possible, both against 
high-redshift radio-loud AGN (e.g Wiklind \& Combes 1998; Murphy, 
Curran \& Webb 2003), or 
distant transient GRB 
sources (see Piro  et al.  2002). 
While there is hope that bright radio 
galaxies can be found at $z>5$ (Jarvis  et al.  2001; Furlanetto 
\& Loeb 2002), 
transients would have the advantage of an effectively
unlimited redshift, and thus the presence of CO absorption at 
rest 115-GHz along a large number of independent lines of sight 
to $z \simeq 20$ could provide 
direct measurements of 
the earliest epoch at which etals became present in the 
ISM of galaxies. In principle, sufficiently bright 
illuminating sources could be used to detect absorption due to a 
wide variety of more complex 
molecules. The redshift of the first molecules is likely to be 
the same as the epoch of the first illuminating beacons, and so it 
presents the opportunity to make a consistency check of the first 
star-formation activity. 

Transient radio pulses should last for hours to days in their 
restframes, 
and then be time dilated to arrive as signals with durations of 
days to months, and so the SKA could in 
principle detect absorbers  
without requiring very rapid response (Furlanetto \& Loeb 2002). 
At 100\,MHz, the instantaneous field of 
view is large, and so in a deep image the possibility of monitoring 
a constantly  
varying set of faint transients can be forseen. The imaging 
quality from long baselines to avoid source confusion is an 
important issue. Arcsec resolution, and 1000\,km baselines would be required.

\subsection{Radio recombination lines (RRLs)} 

RRLs have been sought observationally for 
30 years, but have only 
rarely been useful for revealing practical extragalactic astrophysics 
(see Mohan, Anantharamaiah, \& Goss 2002), owing to 
their low intrinsic luminosities (see Oh \& Mack 2003). 
The sensitivity of the 
SKA offers to change this situation dramatically. 
Maser modification of the line intensities, 
which can only increase their detectability, is likely for transitions at 
lower frequencies than the H50$\alpha$ line at 51.1\,GHz. Note that 
at higher frequencies, 
the lines H$n\alpha$,\footnote{This transition is between the $n+1$ and $n$ 
quantum states, at a frequency close to $3.289 \times 10^{15} [n^{-2} - 
(1+n)^{-2}$\,Hz (Lyman-$\alpha$ is H1$\alpha$ in the same 
terminology} and H$(n+1)\alpha$ are separared by a 
difference in frequency 
that can be calculated using the formula in the footnote, and is of order 
3\,GHz at 50\,GHz.  
In the calculations shown in Fig.\,2 the 
H50$\alpha$ line is assumed to include $1.9 \times 10^{-9}$ 
of the total power emitted by the galaxy (Seaquist, Kerton \& Bell 1994), 
and other lines are assumed 
to be populated thermally, with luminosities proportional to $\nu^2$. 
The frequency separation of adjacent RRLs is of order 6\% of the line 
frequency, and so at least 
2 and usually 3 lines should be present in any 4-GHz window from 
high redshifts, even out to 40\,GHz. Hence, although faint, RRLs could 
provide a direct, single-shot 
redshift indicator for any star-forming galaxies 
detected with enough power to have a signal-to-noise ratio 
greater than about 100 in the radio continuum over the typical
100\,km\,s$^{-1}$ line width of the galaxy. This corresponds 
to a continuum flux density of about 0.4\,mJy 
in a 24-hour integration at a frequency of  
1.4-GHz, or to an ultraluminous galaxy at $z \simeq 2$. 
By detecting 
multiple lines, the exact set of transitions ($n$) 
being observed can be 
determined without any ambiguity, leading to a very accurate redshift for  
the source, with a centroid of $\pm 30$\,km\,s$^{-1}$, indicating a 
redshift accuracy of only $\Delta z 
\simeq 0.0003$ at $z \simeq 2$. 
The relative intensities of the lines can then be used to check 
on possible maser activity and the physical conditions within the 
warm almost-ionized, recombining gas in the source. The optically 
thin emission from RRLs should be related accurately to the 
ionization rate in the ISM 
and/or intergalactic gas around the galaxy, while 
the associated neutral atomic and molecular gas could be probed by 
H{\sc I} and CO(1-0) emission in sufficiently deep observations using 
ALMA or SKA once the redshift was known. 
Taken together, over fields that are many tens of arcmin in extent, 
the SKA should be able to 
trace the history of processing and enrichment of gas in resolved 
detail in RRLs 
over the most active epochs of galaxy formation at $z \simeq 2$--3. 

\subsection{H{\sc i} emission} 

The emission of 21-cm H hyperfine structure radiation is discussed more 
extensively elsewhere (van der Hulst  et al.  2004; 
this volume), 
but it will provide a useful 
way to probe the mass of neutral gas in distant galaxies. With 
a sufficiently long baseline and integration time, 
it may be possible to resolve the internal 
spatial structure of galaxies,  
allowing accurate modeling of the mass distribution within the 
inner baryonic 
part of both primordial gas clouds and 
metal-rich galaxies at high redshifts. Existing facilities can detect the most 
massive objects in H{\sc i} emission to only moderate redshifts (e.g. 
Zwaan  et al.  2001), 
and in absorption in more 
luminous distant example (De Breuck  et al.  2003). A far more sensitive 
telescope, with a wide spectral band could use redshifted H{\sc i} to
trace reionization (Furlanetto \& Loeb 2002), and image 
with sufficient 
spatial resolution to probe galaxies' LSS at $z \simeq 3$ 
over a representative volume, targeting a 5-deg-wide 
swath of sky simultaneously with a 10-m antenna size. 

\subsection{Masers}

Rich series of transitions in OH and H$_2$O molecules 
can be driven into population inversions to generate 
powerful maser emission (Darling, Goldsmith \& Giovanelli 
2002; Darling \& Giovanelli 
2003). Especially important for observations out to high redshifts 
are likely to be 
the 1665/1667-MHz OH maser lines and 1612/1720-MHz 
conjugate lines (Darling 2004). Any telescope that is efficient 
at detecting H{\sc i} emission from low redshifts, can also 
detect these lines, and a very sensitive interferometer like SKA should 
be able to detect them out to high redshift (Darling et al.\ 2004). The 
intense illumination required to drive the gas into a population 
inversion could be readily produced by dense knots of star 
formation activity in even the most distant systems, and so 
the 0.1-1\,GHz band could be very promising for selecting and 
measuring the redshifts of distant galaxies. This is a regime 
where the lower frequency coverage of the SKA can make 
significant contributions, with a modest requirement on 
correlator bandwidth as compared with the demands of a 
CO-based system (Townsend  et al.  2001). The expectation that 
a 2-line maser spectrum would be seen
prevents confusion with intervening lower-redshift H{\sc i} emission, 
and provides direct confirmation of the redshift identification at a 
single frequency  
(Briggs 2000).  

An added, speculative use of the physically simple, small regions 
where maser emission arises is that the relative frequencies 
of different maser transitions can be measured very accurately, 
and are sensitive to any changes in the value of the fine structure 
constant $\alpha$ out to moderate and high redshifts 
(Darling 2003). Highly controversial remains possible evolution of 
the fine-structure constant has been 
suggested based on optical observations of 
QSO absorption lines (e.g. Murphy, Webb \& Flambaum 2003; Chand et al. 
2004). OH 
absorption measures and weak maser lines offer to probe this 
evolution very accurately. The conjugate maser strength is 1\% 
of the continuum, and can be detected out to the maximum 
redshift of bright background galaxies (Darling 2004).  

Maser lines  
provide alternatives to another radio method
using 1.4-GHz H{\sc i} radiation (Carilli  et al.  2000; 
Carilli \& Taylor 2000), and 
one which may be less prone to systematic effects (see Curran  et al.  
2004; this volume). 

The relative spatial positions  
of any OH maser emission within a 200/$(1+z)$\,MHz bandwidth of the SKA
when combined with the frequency of the maser line
could provide a 
very valuable tracer of the kinematics of the maser emission spots with 
a galaxy, and could possibly be used to trace accelerations in multi-epoch 
observations. 

\section{Continuum emission} 

At present the VLA is able to detect the 
synchrotron continuum emission 
from the combined supernovae remnants 
in galaxies out to $z \simeq 3$, but only from those galaxies with 
star-formation rates in excess of several 100\,M$_\odot$\,yr$^{-1}$ 
(Chapman  et al.  2002). The necessary surveys reach 1-$\sigma$ sensitivities 
of several $\mu$Jy. Most have been conducted at 1.4\,GHz, although 
deep 5- and 8-GHz observations have been made; multiband radio 
data is valuable, as a flatter or steeper spectral index than $f_\nu 
\propto \nu^\alpha$ with $\alpha \simeq -0.6$ could be used to spot 
faint AGN amongst the sample of high-redshift star-forming galaxies. 
Measurement of the continuum slope could also provide additional 
astrophysical information (van der Hulst et al.  2004; this volume). 

The emission from synchrotron sources is expected to be reduced 
at higher redshifts, as the intensity of the CMB radiation 
increases as $(1+z)^4$, and provides a loss mechanism by 
scattering the relativistic electrons responsible for powering 
synchrotron emission, and effect that becomes 
important when the energy density in the CMB ($U_{\rm CMB}$) 
matches that in the 
local ISM magnetic field $U_{\rm B}$. This is likely to occur for quiescent 
spiral galaxies until $z \simeq 1$; however, at higher redshifts 
inverse Compton cooling will dominate
synchrotron radiation as a mechanism to remove energy from relativistic 
electrons, leading to shorter 
lifetimes, and a departure to the radio-quiet side of the tight 
radio--far-IR correlation (Condon et al.\ 1992) 

This is not true however for the most luminous 
compact nuclear starburst galaxies with star formation rates 
$>100$\,M$_\odot$ 
yr$^{-1}$ in regions only several 100\,pc in extent 
(Downes \& Solomon 1998). 
In these systems the
energy density in the magnetic field is almost three orders of
magnitude larger than that in the Milky Way disk 
(Carilli \& Taylor 2000), in which
case $U_{\rm CMB}$ only becomes relevant at $z > 7$. However, compact
nuclear starbursts raise a different, related problem: the energy
density in the IR radiation field (U$_\gamma$) from the starburst
itself is larger than that in the magnetic field. This means
that inverse Compton cooling should remove the synchrotron emitting
electrons on fairly short timescales ($\le 10^5$\,yr), and accentuates
the question: why do low-redshift nuclear starbursts follow the 
radio--far-IR
correlation?  There is a large body of literature on this issue,
but as yet no resolution. Hence, we trade one problem --
inverse Compton losses off the microwave background at high redshift -- 
for a
second -- inverse Compton losses off the starburst infrared radiation field at
all redshifts.

The densest, most energetic regions 
of star formation may be able to sustain synchrotron emission 
to higher redshifts, but during the extremely distant epoch of 
reionization at $z \simeq 6$-20, only short-lived 
GRB and AGN jets will be detectable in synchrotron emission.  

Free-free continuum emission will always remain detectable, however, 
whenever 
high-mass star formation is taking place. While the intensity of 
nebular free-free emission is less than that of synchrotron emission in 
nearby galaxies (Condon 1992), 
a sufficiently sensitive telescope 
could detect the associated continuum sources out to $z \simeq 20$. 
Excellent angular resolution would be required to identify 
individual sources. 

Fig.\,1 shows the expected count of very faint star-forming galaxies 
and radio-quiet AGN expected at a variety of frequencies from 
0.5 to 8\,GHz, including the expected effect of CMB emission 
suppression, with and without the contribution of 
free-free emission, based on models 
described in more detail by 
Blain  et al.  (2002) and Carilli \& Blain (2002). 
The effect of free-free emission 
at boosting the number of faint, high-redshift galaxies is clear, 
especially when observing at the higher SKA frequencies. 

Note that it will never be possible to detect thermal 
dust emission from very high 
redshift galaxies using cm-wave telescopes, as the dust temperature 
cannot fall below that of the CMB. At the highest redshifts ($z > 10$),  
the K-correction that is so beneficial for detecting distant galaxies at 
submm wavelengths (Blain \& Longair 1993; 
Blain  et al.  2002) is prevented from operating 
further, and free-free emission is expected to be the most powerful source of 
radiation at all restframe frequencies of order 100-300\,GHz that will 
be probed by SKA at the highest redshifts (Blain  et al.  2002). 

\section{Gravitational lensing} 
 
With the potential to resolve the shapes of high-redshift galaxies, 
and to determine their redshifts using RRLs over a large field in a single 
very deep exposure, it could be possible to use SKA in deep integrations 
to generate a cosmic shear map that highlights the mass in foreground 
structures. The 0.1-arcsec spatial resolution is necessary to resolve 
high-redshift galaxies, and is thus essential for this type of 
study. Without knowing the nature of the morphology and internal 
spatial distribution 
for the emission from numerous 
distant radio sources in detail it is difficult to assess the true 
capabilities (Schneider  et al.  2000; Blake et al.  2004, this volume). 
However, the chances of 
success are very high, as SKA imaging quality and depth 
available should be sufficient. Both subarcsec resolution and $\mu$Jy 
sensitivities are absolute requirements for this goal, while a 
wide instantaneous field of view 
would be very useful to generate a sufficiently 
large number ($\simeq 10^5$) of detected galaxies with accurate 
shapes (Fig.\,1). 

The most massive foreground galaxies, ellipticals out 
to $z \simeq 1$, will be abundant in every SKA pointing. In 
their vicinity will be 
the faint radio images  
of background galaxies distorted and magnified 
by the ellipticals. These images will appear in all SKA pointings that 
are sufficiently deep ($\simeq 10$\,nJy RMS) to have a source density  
of order 10$^6$\,deg$^{-2}$ 
No dedicated observing time will be necessary for these observations. 
These massive, non-star-forming 
foreground galaxies are not expected to be strong radio sources, with 
the possible exception of point sources in their nuclei. and so the 
contrast between lens and image should be very high, in contrast 
to optical studies of galaxy--galaxy lensing for which the emission 
from the foreground lens dominates. 
The images whose morphologies 
indicate the strongest distortion and thus the highest 
magnification would be excellent candidates
for resolved studies one by one using ALMA. 

\section{Conclusions} 

Wide-field wideband spectra at radio wavelengths offer several 
unique opportunities for studying the evolution of both galaxies 
themselves 
and their large-scale structure. Radio recombination lines at 
low redshifts, highly redshifted molecular lines from close to the 
end of recombination, and absorption lines against sources prior 
to the end of recombination are all previously barely explored 
phenomena that will be relatively easy to investigate using an 
SKA. The possibilities for its exploitation of these effects 
would advocate the emphasis of the 
following instrument design features: 

\begin{enumerate} 
\item For all surveys of high-redshift galaxies, resolutions of 
1\,arcsec are essential to avoid confusion, and ideally resolutions 
better than 0.1\,arcsec would allow substructure to be probed within 
the distant objects, imposing a minimum baseline requirement of 
150\,km for frequencies as high as 40\,GHz, and an Earth-sized
baseline at the lowest frequencies. 
\item To detect low-excitation CO and HCN emission from cold molecular 
gas in distant galaxies it is necessary to push the frequency 
range of 
the SKA up towards the 40\,GHz atmospheric cutoff, to ensure that the 
radiation from the most active phase of galaxy evolution at $z \simeq 2$--3 
can be probed. 
\item For LSS surveys, the most powerful, most 
easily reconfigurable wide-band correlator is required. The ability to 
simultaneously correlate a spectrum over a large fraction of the 
receiver bandwidth would be a very useful goal, providing the 
capability for
line surveys with great depth in redshift. 
\item For all deep surveys it is more desirable to have a larger number 
of smaller antennae for the same total collecting area in order 
to maximize the instantaneous field of view. This 
would maximize the efficiency of the telescope, and highlight some of the 
faintest yet most interesting galaxies for study using ALMA, {\it 
JWST} and future 
30-m optical/near-IR ground-based telescopes. 
\end{enumerate} 

To summarize, 
an instrument 
with 10$^4$ 10-m apertures, a maximum baseline of 1000\,km, and a very 
powerful correlator would be extremely valuable for studying galaxy 
evolution at radio wavelengths. 

\onecolumn
\begin{figure}
\begin{center}
\psfig{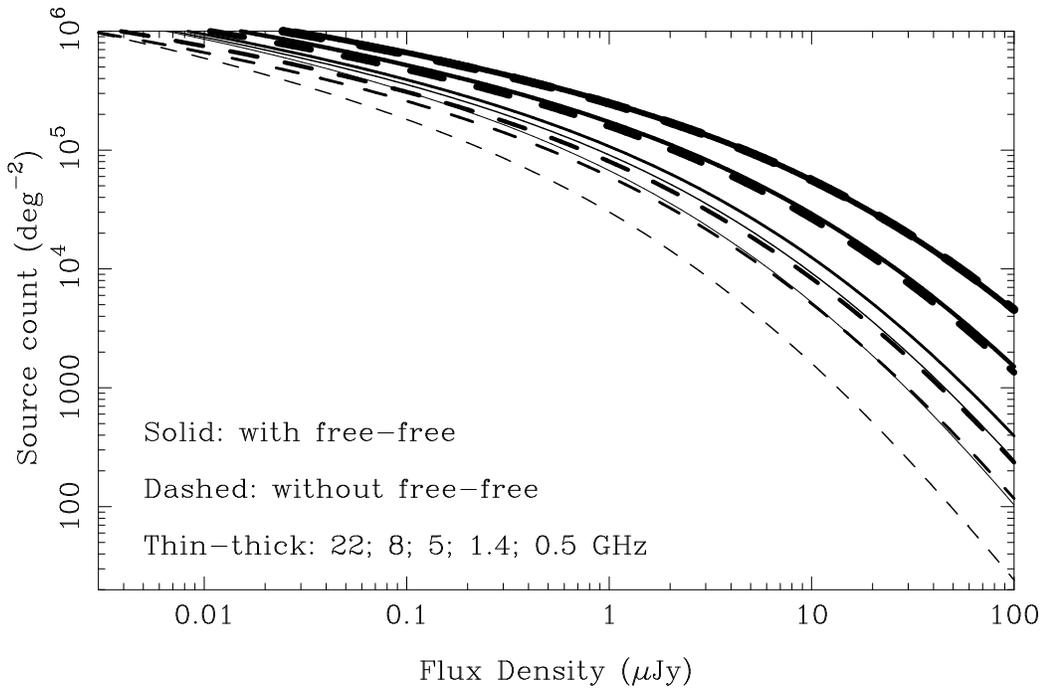}
\end{center} 
\vspace{-0.2in}
\caption{The integral count $N(>S)$ 
of continuum galaxies expected at extremely faint radio 
flux density levels $S$, for several different SKA wavelengths, 
based on a model in which the radio population is extrapolated from 
dusty star-forming galaxies (see Blain et al.  2002). 
Existing surveys reach detection limits no fainter than 
10\,$\mu$Jy at 1.4\,GHz.
The maximum flux is below the cutoff at which the count of radio-loud 
AGN falls beneath that of star-forming galaxies in existing 1.4-GHz 
surveys. There should be a \emph{very} large sample of continuum-emitting 
galaxies for the 
SKA to detect. 
}
\end{figure}

\begin{figure}
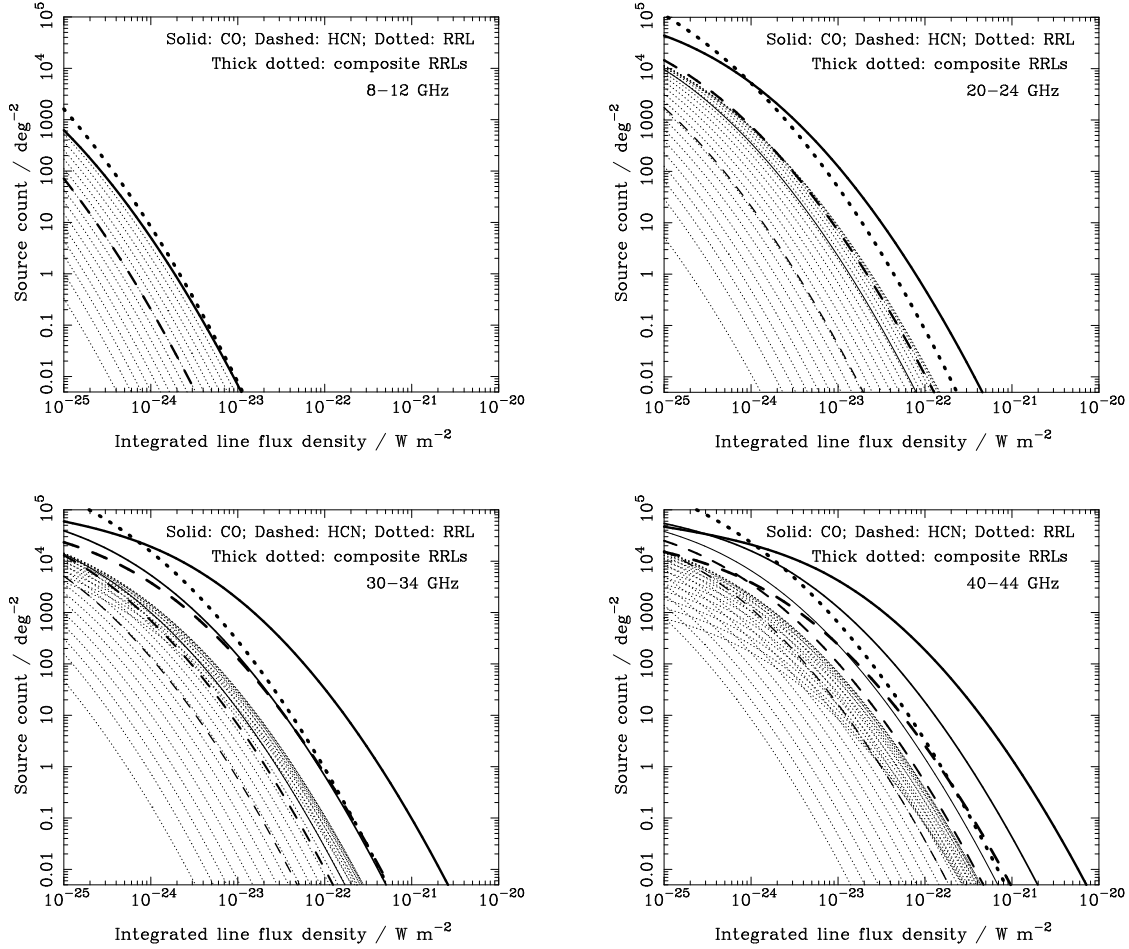

\begin{center} 
\psfig{figure=line0812.ps,width=2.7in,angle=-90} \vspace{-6.0cm} \hspace{6.7cm} 
\psfig{figure=line2024.ps,width=2.7in,angle=-90} 
\end{center} 
\begin{center} 
\vspace{0.2cm} 
\psfig{figure=line3034.ps,width=2.7in,angle=-90} \vspace{-6cm} \hspace{6.7cm} 
\psfig{figure=line4044.ps,width=2.7in,angle=-90}
\end{center} 
\caption{The count of line-emitting galaxies expected out to redshift 
$z=10$ in four higher-frequency 
potential SKA bands, each 4\,GHz wide. The thickest solid and 
dashed lines represent the (1-0) transitions of CO and HCN respectively; 
progressively thinner lines of the same style represent higher $J$ transitions
($J+1\rightarrow J$). In the lower frequency bands the (2-1) and (3-2) 
transitions are not redshifted far enough to be detectable. 
The higher-excitation RRLs 
at higher frequencies than $n=50$ (51.1\,GHz) 
are shown (dotted lines), In each band, the dotted lines 
represent lines in the RRL ladder with lower $n$ values, at the 
lower left side, starting with $n=19$: dotted lines with progressively 
greater values of $n$ 
appear at the upper right, until the count is maximized for a 
certain value of $n$, and the count then declines (as shown by the 
flat low-flux counts for 40-44\,GHz). At lower 
excitations ($n>50$), the effect of masing could boost the RRL line strength, 
and so the contribution of RRLs could be greater than these 
results suggest.  
Note that in 
the combined total count of galaxies, RRLs (at lower 
redshifts than the CO and HCN sources) could contribute a comparable 
number of detections as the molecular lines, and so could provide a 
very valuable redshift survey at moderate redshifts, especially $z \simeq 1.5$ 
for the 20-24\,GHz band.  
In a 1-hr integration 
at 42\,GHz, a 5-$\sigma$ sensitivity of about 
$1.2 \times 10^{-24}$\,W\,m$^{-2}$ is expected for a  
300-km-s$^{-1}$ wide line, while at 22\,GHz the 
corresponding value is $2.8 \times 10^{-24}$\,W\,m$^{-2}$. This corresponds 
to of order 50 line detections within the 0.03-deg-wide 
primary beam of a 12-m element SKA in a 1-hr 
integration, and thus to a viable redshift survey.}
\end{figure}
\twocolumn

\vskip 0.2truein 

AWB acknowledges support from NSF grant AST-0205937, from the 
Research Corporation and from the Alfred P. Sloan Foundation.  We thank 
Joy Chavez for compiling data on RRLs during a summer internship at 
Caltech in 2003, supported by the Caltech Minorities Undergraduate 
Research Fellowship program (MURF).

\end{document}